\documentclass[a4paper,11pt]{article}

\usepackage[T1]{fontenc}
\usepackage[utf8]{inputenc}
\usepackage{graphicx}
\usepackage{xcolor}
\usepackage{float}
\usepackage{caption}

\usepackage{tgtermes}
\usepackage{amsmath,amssymb,amsthm,textcomp}
\usepackage{enumerate}
\usepackage{multicol}
\usepackage{tikz}
\usepackage{authblk}

\usepackage{geometry}
\newtheoremstyle{mytheor}
    {1ex}{1ex}{\normalfont}{0pt}{\scshape}{.}{1ex}
    {{\thmname{#1 }}{\thmnumber{#2}}{\thmnote{ (#3)}}}

\theoremstyle{mytheor}

\makeatletter
\renewcommand{\maketitle}{
\begin{center}
\vspace{2ex}
{\huge \textsc{\@title}}
\\
\rule{\textwidth}{1.5pt}\\ 
 \@author \hfill  
\vspace{4ex}
\end{center}
}
\makeatother

\usepackage{fancyhdr,lastpage}
\title{\Large Mobility restrictions for the control of epidemics: When do they work?}

\author[a,c]{Baltazar Espinoza}
\author[a,c]{Carlos Castillo-Chavez} 
\author[b]{Charles Perrings}

\affil[a]{Division of Applied Mathematics, Brown University, Providence, RI 02903.}

\affil[b]{School of Life Sciences and ecoSERVICES Group, Arizona State University, Tempe, AZ 85287-4501}
\affil[c]{Simon A. Levin Mathematical and Computational Modeling Sciences Center, Arizona State University, Tempe, AZ 85287-3901}

\begin{document}

\maketitle
\vspace{-1cm}

\begin{abstract}
Mobility restrictions - travel advisories, trade and travel bans, border closures and, in extreme cases, area quarantines or \emph{cordons sanitaires} - are among the most widely used measures to control infectious diseases. Restrictions of this kind were important in the response to epidemics of SARS (2003), H1N1 influenza (2009), and Ebola (2014). However, they do not always work as expected. The imposition of a \emph{cordon sanitaire} to control the 2014 West African Ebola outbreak, for example, is argued to have led to a higher-than-expected number of cases in the quarantined area. To determine when mobility restrictions reduce the size of an epidemic, we use a model of disease transmission within and between economically heterogeneous locally connected communities. One community comprises a low-risk, resource-rich, low-density population with access to effective medical resources. The other comprises a high-risk, resource-poor, high-density population without access to effective medical resources. We find that the overall size of an epidemic centered in the high-risk community is sensitive to the stringency of mobility restrictions between the two communities. Unrestricted mobility between the two risk communities increases the number of secondary cases in the low-risk community but reduces the overall epidemic size. By contrast, the imposition of a \emph{cordon sanitaire} around the high-risk community reduces the number of secondary infections in the low-risk community but increases the overall epidemic size. The degree to which mobility restrictions increase or decrease the overall epidemic size depends on the level of risk in each community and the characteristics of the disease.
\end{abstract}

\section{Introduction}
The 2003 Severe Acute Respiratory Syndrome (SARS) epidemic, the 2009 influenza A (H1N1) pandemic, and the 2014 West African Ebola Virus Disease (EVD) epidemic provide recent reminders that the rapidity and extent of the spread of infectious disease depends on patterns of human mobility. Pre-existing patterns of trade and travel determine the routes along which diseases may potentially spread \cite{hufnagel2004forecast}. While mobility may change once an outbreak is known to have occurred if people choose to mitigate the risk to themselves \cite{fenichel2013skip, perrings2014merging}, the underlying patterns of human mobility set the baseline against which health authorities decide when and where to impose travel restrictions, to close borders or, in extreme cases, to establish area quarantines -- \emph{cordons sanitaires}.

Mobility restrictions have a long history. Examples of the use of \emph{cordons sanitaires} include measures to stop the bubonic plague (1666) \cite{race1995some, wallis2006dreadful}, yellow fever (1793, 1821, 1882) \cite{arnebeck2008short, kohn2007encyclopedia}, and cholera (1830, 1884) \cite{kohn2007encyclopedia}. In many such cases, restrictions have involved the deployment of physical barriers secured by armed forces. The implementation of such measures can then be seen as infringing the rights of people, as well as being both cumbersome and expensive \cite{cetron2005public, ebolaethic, NYT2014ebola, cdc_ethic, cnn_ethic}. In some cases, mobility restrictions for disease control have had catastrophic consequences for the affected population \cite{nigeria_senegal, cordon_escape}.

Nor are mobility restrictions always successful in the control of the disease \cite{davis2013spanish, towers2014temporal}. One example is the \emph{cordon sanitaire} implemented during the 2014 Ebola Virus Disease (EVD) epidemic in West Africa, in which 28,600 cases resulted in more than 11,000 deaths, \cite{CDC_count, kuhn2010proposal, gire2014genomic}. The \emph{cordon sanitaire} was applied to the area containing, at the time, more than 70\% of the epidemic in an effort to contain the spread of the disease \cite{cordons}. Travel restrictions produced a humanitarian crisis within the quarantined region. Disruption of the food transportation system led to food shortages, while lack of appropriate health care increased the risk of infection \cite{cordon_polemic, ebolaethic}. Lack of mobility and growing levels of infection resulted in an increasing (effective) reproduction number over time, and therefore also in the number of EVD cases. Indeed, the data suggest that mobility restrictions may have accelerated the contagion process, and therefore led to a higher than expected number of cases within the mobility-regulated region \cite{towers2014temporal, espinoza2016assessing, Pandey2014}.

The question raised by such cases is when are mobility restrictions an effective means of disease control in neighboring communities? The impact of travel restrictions on disease risk has been previously analyzed both theoretically (largely from an Eulerian perspective on mobility \cite{colizza2008epidemic, arino2007quarantine}), and empirically (using, for example, data on the impact of international air travel restrictions on disease spread during the 2009 Swine flu pandemic \cite{epstein2007controlling, bajardi2011human} and the 2014 Ebola outbreak \cite{ poletto2014assessing, gomes2014assessing}). Most such studies have focused on the role of restrictions on the rate of spread rather than the final epidemic size, and have shown that travel restrictions can slow the rate at which a disease spreads from the source of infection.
While the speed of transmission is relevant, our primary concern is to understand the implications of mobility for the location of secondary infections relative to the availability and quality of health care. That is, to quantify the role of local mobility patterns on the local and total number of secondary cases, the final epidemic size.

In this paper we develop a two communities model in which the communities are connected through the movement of people. 
We model the time spent by individuals within each community, and use this to study the dynamics of infectious disease under multiple mobility regimes. The model keeps track of individuals’ place of residency through the incorporation, via a residency time matrix, of the proportion of time spent by the resident in his/her own community (a Lagrangian approach). The balance of the time available is assumed to be spent as a visitor to the second community.
The communities are differentiated by socio-economic conditions such as income, wealth, public health infrastructure, and risk.
Those in one community face a high-risk of infection, those in the other face a low-risk of infection.
We then use the model to assess the effectiveness of disease control via mobility restrictions. We use EVD as our model disease, but note that our results hold for a range of disease types.

Our results suggest that tight mobility restrictions can increase the overall level of infection. Weak mobility restrictions can have the opposite effect. We test the effect of differing mobility levels on the final epidemic size, and find that both ``intermediate'' and ``high'' mobility levels may reduce the overall final epidemic size. Indeed, in extreme cases removing all restrictions on mobility levels may be sufficient to control an outbreak. We identify two mobility thresholds. One is the level of mobility needed to do better than the \emph{cordon sanitaire}. The second is the level of mobility needed to control the disease. Since population density play a role in disease dynamics, we test the sensitivity of our findings to variation in population density ratios, and to community-specific risks of infection.

\section{Disease Dynamics in Heterogeneous Communities: a Lagrangian approach}

We employ a Lagrangian approach, which uses a \emph{residency time matrix} that tracks individuals' mobility between two communities. The model incorporates the average proportion of time that individuals spend in each community as elements of the matrix $\mathbb{P}=(p_{ij}),\; i,j\in \{1,2\}$, where $p_{ij}\geq 0$ are assumed to be constant over time. The Lagrangian perspective allows us to assess the impact of population mobility on overall disease dynamics, \cite{bichara2015sis, espinoza2016assessing, castillo2016perspectives}.

The global population of interest is assumed to comprise two linked communities facing distinct levels of EVD infection risk.  Differences in infection risk are captured by a single parameter ($\beta_i$) acting on the community effective population size. The risk of infection reflects community attributes that include income, education, health-care access, cultural practices, and so on.
In the absence of mobility, Community $1$ is assumed to be capable of sustaining an epidemic ($\mathcal{R}_{01}>1$) while Community $2$ is assumed to be unable to  support an outbreak in isolation, ($\mathcal{R}_{02}<1$). The model is calibrated using data from the West African EVD outbreak, which gives a value of $\mathcal{R}_{01}=2.45$, \cite{towers2014temporal, chowell2014transmission, althaus2014estimating}. 
Detailed model formulation, computation of the community-specific and global basic reproductive numbers obtained using the next generation approach \cite{MR1057044,VddWat02}, as well as community-specific and global final epidemic size, can be found in the SI appendix.

\section{Disease Control Through Mobility Restrictions}
We explore the impact of mobility from the high-risk community (HRC) and low-risk community (LRC) on the final epidemic size. In particular, we consider the conditions under which a \emph{cordon sanitaire} is effective, focusing on the conditioning effects of population densities and risk disparities.

\subsection{The Impact of High-Risk Community Mobility}
Intuitively, one would expect that movement of infected and infectious individuals to a region consisting only of susceptible individuals would increase the overall final epidemic size. However, if the infected individuals move to a region having better sanitary conditions, an increase in the number of secondary infections in the LRC may be offset by a reduction in the number of secondary infections in the HRC.

\begin{figure}[tbhp]
\centering
\includegraphics[width=0.8\linewidth]{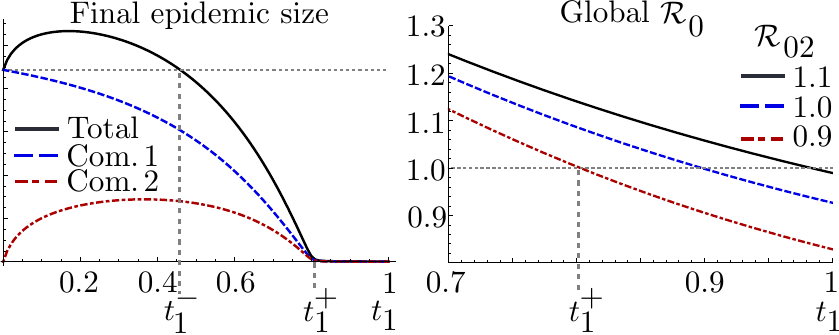}
\caption{(Left panel) Community specific and total final epidemic size under one way mobility $(t_2=0)$. (Right panel) Global $\mathcal{R}_0$ for different Community $2$ risk scenarios, $t_2=0$.}
\label{fig:threefinals}
\end{figure}

Figure \ref{fig:threefinals} (left panel) depicts the community specific and combined final epidemic sizes as a function of the average proportion of time that Community $1$ residents spend in Community $2$, ($t_1$).
Individuals from the safer community are assumed to avoid the HRC. That is, our two communities model is calibrated under the assumption that $t_2=0$.
Taking the final epidemic size corresponding to the \emph{cordon sanitaire} scenario as a baseline ($t_1=0$, dashed gray line), and assuming that $\mathcal{R}_{02}=0.9$, we see that low mobility levels ($t_1<0.45$) increase the total final epidemic size relative to the baseline, but that moderate mobility levels ($t_1>0.45$) reduce the total final epidemic size below the \emph{cordon sanitaire} scenario. Moreover, ``high'' levels of single direction mobility ($t_1>0.8$) lead to the control of an ongoing EVD outbreak. This indicates the existence of a sharp threshold disease persistence condition in the two communities system operating under a Lagrangian mobility framework modeled via residence times, \cite{bichara2015sis}.

We identify two thresholds: the levels of mobility required to reduce the total final size below the \emph{cordon sanitaire} scenario ($t_1^-$), and the levels of mobility needed to control an EVD outbreak in the whole system ($t_1^+$). The second empirical threshold determines the levels of mobility needed for the global, integrated communities, basic reproductive number ($\mathcal{R}_0(\mathbb{P})$) to fall below $1$. Particularly, $t_1^-$ is seen to capture the trade off between diminishing cases in Community $1$ and increases in the number of infected individuals in Community $2$.

Figure \ref{fig:threefinals} (right panel) shows the threshold beyond which unidirectional mobility can control an EVD outbreak, as a function of Community $2$ risk of infection. The scenario highlighted in Figure \ref{fig:threefinals} (left panel) corresponds to $\mathcal{R}_{01}=2.45$ and $\mathcal{R}_{02}=0.9$. We see that mobility above $t_1=0.8$, supports a global $\mathcal{R}_0$ less than one, leading to a final epidemic size near zero. It is worth observing that the curves in Figure \ref{fig:threefinals} (right panel) do not converge to $\mathcal{R}_{02}$ at the extreme value $t_1=1$, this is because our two communities model is asymmetric (due, for example, to the local management of EVD-infected corpses).

\begin{figure}[tbhp]
\centering
\includegraphics[width=0.8\linewidth]{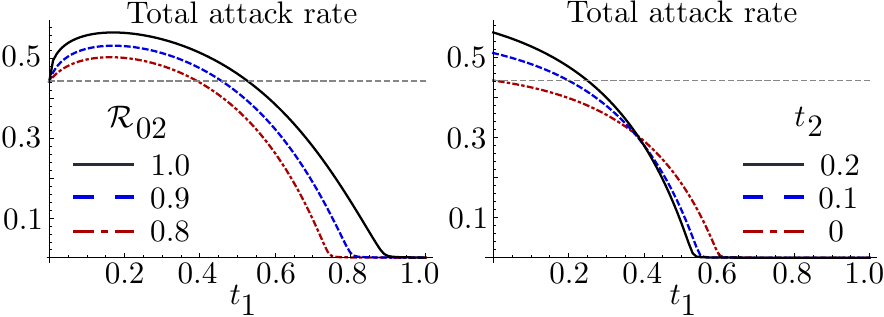}
\caption{(Left panel) Total attack rate for different Community $2$ risk levels, under one-way mobility and $\mathcal{R}_{01}=2.45$. (Right panel) Attack rates under two-ways mobility, $\mathcal{R}_{01}=2.45$ and $\mathcal{R}_{02}=0.1$.}
\label{fig:PNAS_threeattack}
\end{figure}

Figure \ref{fig:PNAS_threeattack} (left panel) shows that both empirical thresholds described by $t_1^-$ and $t_1^+$  are highly sensitive to the risk of infection in Community $2$, Figure \ref{fig:PNAS_threeattack} (right panel) shows that low mobility from a highly safe Community $2$ reduces $t_1^+$. One might therefore conclude that improvements in Community $2$ sanitary conditions plays a dual role: reducing the overall number of EVD secondary infections, and relaxing the mobility conditions required to manage an epidemic on the overall system (if such a policy can be put in place). It also suggests the possibility of using economic incentives to promote appropriate mobility patterns during health emergencies.

\begin{figure}[tbhp]
\centering
\includegraphics[width=0.7\linewidth]{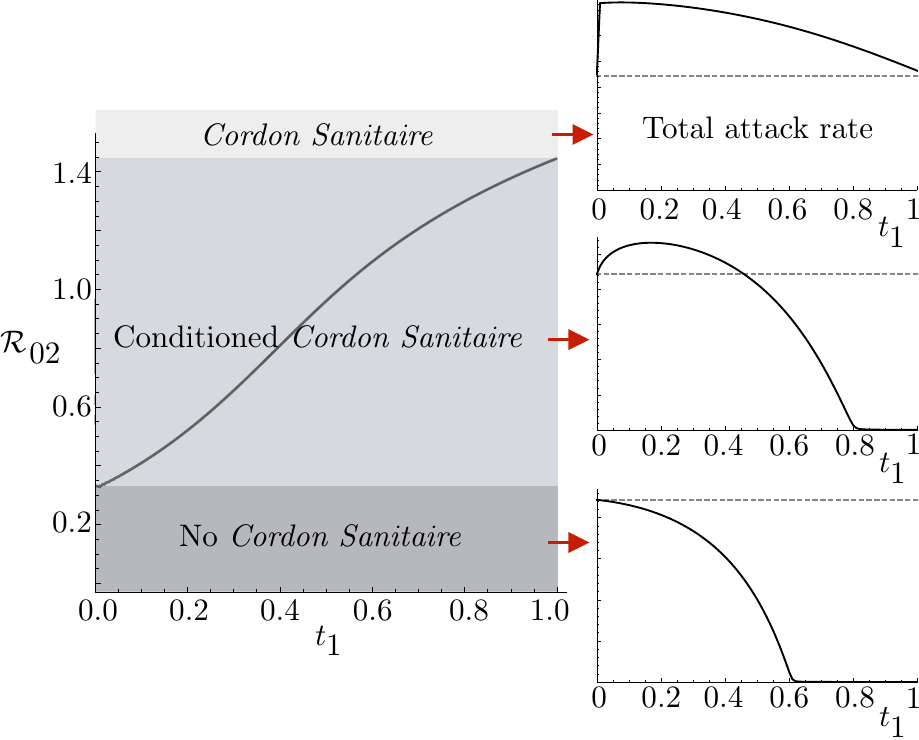}
\caption[\emph{Cordon sanitaire} threshold as function of $\mathcal{R}_{02}$]{Traveling time reduces or increases the total attack rate as function of the Community $2$ risk of infection, ($\mathcal{R}_{01}=2.45$).}
\label{fig:travben}
\end{figure}

Figure \ref{fig:travben} shows the impact of mobility from the HRC on the effectiveness of a \emph{cordon sanitaire} given conditions in Community $2$. The conditions in which the total attack rate increases or decreases with mobility may be summarized as follows:

\begin{itemize}
\item For a ``highly safe'' Community $2$, ($\mathcal{R}_{02} < 0.35$), all mobility levels from the HRC are beneficial. That is, the total attack rate monotonically decreases as $t_1$ increases. Hence, implementation of a \emph{cordon sanitare} under this scenario is the worst possible decision.
\item Given an ``intermediately safe'' Community $2$, ($0.35 < \mathcal{R}_{02} < 1.45$), depending on mobility levels, the total attack rate either increases or decreases. Therefore, under these scenarios, the \emph{cordon sanitaire} is effective provided that mobility levels required to reduce the total attack rate are not attainable. In other words, the \emph{cordon sanitaire} is recommended whenever mobility from the HRC is below $t_1^-(\mathcal{R}_{02})$. Figure \ref{fig:trav_pops} (left panel) shows the mobility levels for which the \emph{cordon sanitaire} is recommended, for same populations densities and $\mathcal{R}_{01}=2.45$.
\item For an ``unsafe'' Community $2$ ($\mathcal{R}_{02} > 1.45$), all mobility levels increase the total attack rate. In these scenarios, even when Community $2$ is considerably safer than Community $1$, the reduced risk of infection is not enough to produce an overall benefit in terms of the total number of infections. Therefore, in these scenarios the \emph{cordon sanitaire} is an effective control strategy.
\end{itemize}

\begin{figure}[tbhp]
\centering
\includegraphics[width=0.8\linewidth]{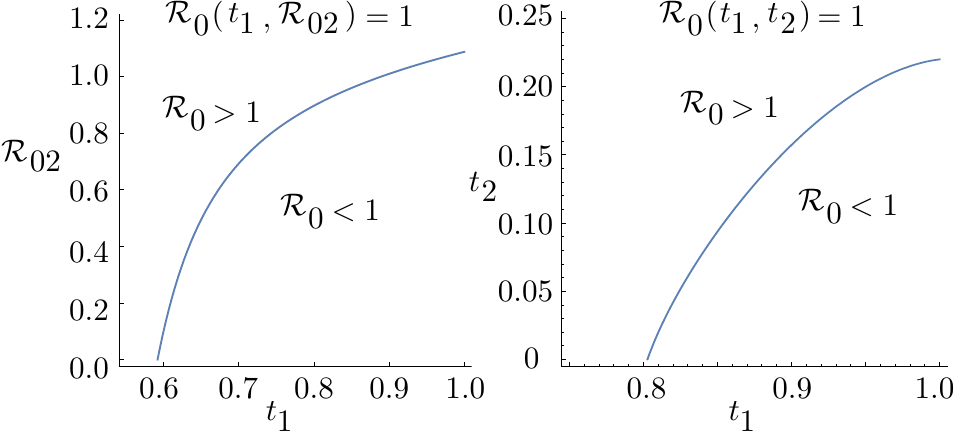}
\caption{(Left panel) Level curve $\mathcal{R}_0(t_1,\mathcal{R}_{02})=1$ in the plane $(t_1,\mathcal{R}_{02})$. Mobility from HRC can eradicate an EVD outbreak, ($\mathcal{R}_{01}=2.45$, $N_1=N_2$). (Right panel) Level curve $\mathcal{R}_0(t_1,t_2)=1$, for $\mathcal{R}_{01}=2.45$ and $\mathcal{R}_{02}=0.9$.}
\label{fig:PNAS_Ro}
\end{figure}

In short, the \emph{cordon sanitaire} does not always reduce the overall number of infected individuals, while our simulations suggest that under specific risk and mobility conditions it might have a detrimental effect. In the simplest scenario of two equally dense communities, the \emph{cordon sanitaire's} effectiveness is determined by the Community $2$ risk of infection. It follows that implementation of a \emph{cordon sanitaire} should depend on the specific attributes of the communities of interest. The state of the health care system in the safer community is critical to the effectiveness of a mobility ban.
The simulations reported in Figure \ref{fig:PNAS_Ro} (left panel) show that mobility, combined with ``good'' enough sanitary conditions in the safe community, may be enough to stop an EVD outbreak. For the model calibrated on data from the 2014 West African Ebola outbreak ($\mathcal{R}_{01}=2.45$), we see that high mobility by itself can lead to a global basic reproductive number below the critical threshold, even when $\mathcal{R}_{02}$ is slightly greater than one. It is important to observe that in the polar case of a completely safe Community $2$ ($\mathcal{R}_{02}=0$), if mobility of the HRC is to control the outbreak, residents of the HRC need to spend at least  60\% of their time in the LRC.

We recognize that high levels of mobility also impose costs. Aside from the cost of treating infected non-residents, some fraction of the LRC residents will become infected. High levels of mobility of infected individuals from the HRC would increase morbidity in the LRC, while reducing morbidity for the integrated communities. Symmetrically, low levels of mobility of infected individuals from the HRC would reduce morbidity in the LRC, while increasing morbidity for the integrated communities. Which outcome is preferred depends on the health authority's objectives.  We return to this issue in the discussion.

\subsection{The Impact of Low-Risk Community Mobility}

The impact of altering two-way mobility levels on the EVD dynamics is studied. We focus on the effect of two-way mobility levels on the \emph{cordon sanitaire} threshold $(t_1^-)$ and elimination threshold $(t_1^+)$.

Figure \ref{fig:PNAS_Ro} (right panel) shows that given a Community $2$ with local basic reproductive number slightly below one ($\mathcal{R}_{02}=0.9$), any increase in mobility from the LRC $(t_2>0)$ has detrimental effects on the overall control of EVD.
That is, increasing $t_2$ demands a greater $t_1^+$ in order to hold $\mathcal{R}_0<1$. Zero mobility from Community $2$ minimizes $t_1^+$.

\begin{figure}
\centering
\includegraphics[width=0.8\linewidth]{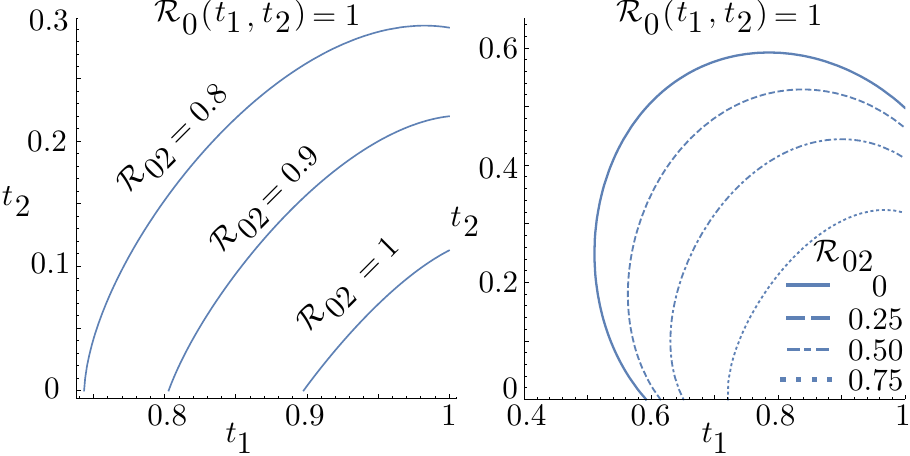}
\caption{(Left panel) Level curves of  $\mathcal{R}_0(t_1,t_2)=1$ for $\mathcal{R}_{01}=2.45$ and $\mathcal{R}_{02}=1,0.9,0.8$. (Right panel) Level curves $\mathcal{R}_0(t_1,t_2)=1$, for $\mathcal{R}_{01}=2.45$ and $\mathcal{R}_{02}=0,0.25,0.5,0.75$.}
\label{fig:PNAS_R0levels}
\end{figure}

In addition, the simulations reported in Figure \ref{fig:PNAS_R0levels} (left panel) show that, for example, when mobility from the LRC rises above $t_2\approx 0.23$ (given that $\mathcal{R}_{02}=0.9$) then there is no mobility from the HRC capable of bringing the global $\mathcal{R}_0$ below one, ($t_1^+$). 
That is, whenever $\mathcal{R}_{02}\approx 1$ and individuals from the LRC are periodically exposed to a high-risk of infection then mobility from the safe community may cancel the beneficial impact of mobility out of the HRC.


In contrast to the case $\mathcal{R}_{02} \approx 1$, mobility from an extremely safe region ($\mathcal{R}_{02}\approx 0$), may help on reducing the final size of an epidemic.
Figure \ref{fig:PNAS_threeattack} (right panel) shows that for $\mathcal{R}_{02}=0.1$, $t_2=0.2$ reduces $t_1^+$, compared to the scenario $t_2=0$; while Figure \ref{fig:PNAS_R0levels} (right panel) shows that a two-ways mobility strategy minimizes $t_1^+$ whenever $\mathcal{R}_{02}\approx 0$.

The benefit obtained by allowing individuals from the LRC to spend time in the HRC can be explained in terms of the link between secondary infections among individuals in the HRC versus secondary infections among individuals in the LRC. The epidemic, in this polarized scenario, is primarily driven by infections among Community $1$ individuals, however LRC mobility reduces the HRC attack rate, without increasing the LRC attack rate significantly.

\subsection{The Effects of Community Density Disparities}

The effects of population density may be addressed by considering the effect of changes in the ratio LRC to HRC residents, combined with the levels of mobility needed to reduce the final epidemic size below the \emph{cordon sanitaire} threshold ($t_1^-$), and the mobility levels required  to stop an ongoing EVD outbreak ($t_2^+$).
 
\begin{figure}[H]
\centering
\includegraphics[width=0.8\linewidth]{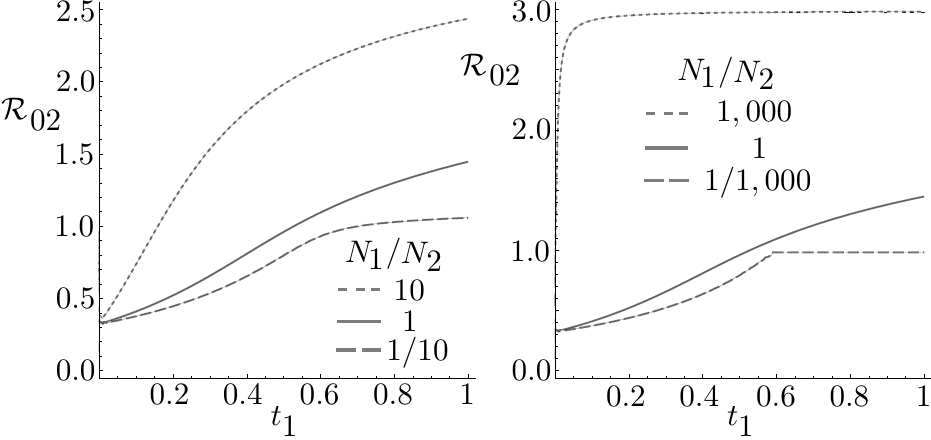}
\caption{(Left panel) \emph{Cordon sanitaire} level curves for population density ratios $\frac{N_1}{N_2}=\frac{1}{10},1,10$. (Right panel) Extreme aggregation scenarios show convergence of mobility thresholds.}
\label{fig:PNAS_cordthresh}
\end{figure}

Figure \ref{fig:PNAS_cordthresh} (left panel) shows the curves of the total attack rate at the \emph{cordon sanitaire} level in the ($t_1,\mathcal{R}_{02}$) plane for the community population density ratios given by, $\frac{N_1}{N_2}=k=10,1,\frac{1}{10}$.
The impact of this ratio is explored when $\mathcal{R}_{01}=2.45$, and $\mathcal{R}_{02}>0.35$  corresponding to the regions where the \emph{cordon sanitaire} conditionally works, which turn out to be sensitive to the ratio of population density in the LRC relative to the HRC. The risk-mobility conditions $\mathcal{R}_{02}<0.35$ correspond to regions where the \emph{cordon sanitaire} is not effective, independent of  population density.

The simulations reported in Figure \ref{fig:PNAS_cordthresh} (left panel) show that as the population density ratio $\frac{N_1}{N_2}$ increases, the minimum mobility level required to drop the total attack rate below the \emph{cordon sanitaire} level also increases. For instance, under the scenario $t_2=0$, $N_1=\frac{1}{10}N_2$ and $\mathcal{R}_{02}=0.5$, the \emph{cordon sanitaire} is recommended provided the mobility $t_1\approx 0.25$ is not attainable. On the other hand, any $t_1>0.25$ leads to a total attack rate below the \emph{cordon sanitaire} level meaning that a \emph{cordon sanitaire} is not recommended at such mobility levels.
In short, a relatively safe population neighboring a smaller unsafe population may benefit from a \emph{cordon sanitaire}, (for parameters $\mathcal{R}_{02}<0.35$). 

\begin{figure}[H]
\centering
\includegraphics[width=0.8\linewidth]{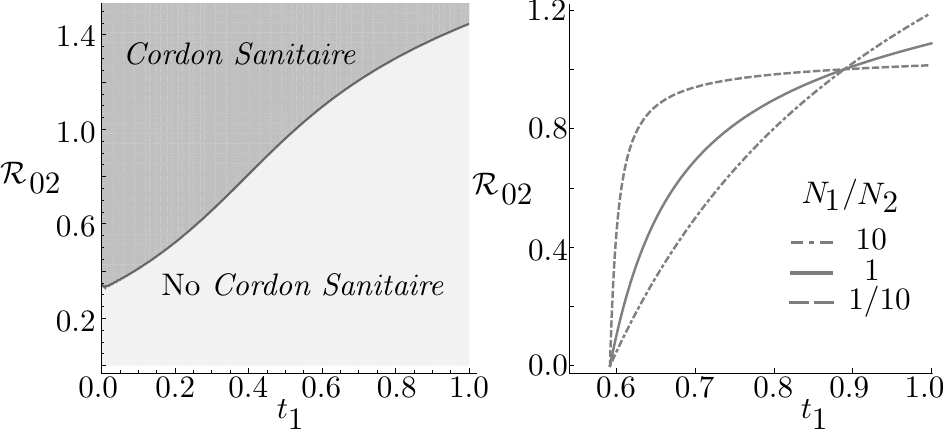}
\caption[Global $\mathcal{R}_0$ varying population densities]{(Left panel) Mobility regions for which the \emph{cordon sanitaire} is effective, assuming  identical populations densities. (Right panel) Level curve $\mathcal{R}_0(t_1,\mathcal{R}_{02})=1$ is for populations ratios $\frac{N_1}{N_2}=\frac{1}{10},1,10$.}
\label{fig:trav_pops}
\end{figure}

In Figure \ref{fig:trav_pops} (right panel) we see how differences in population sizes affect the control of EVD in two communities with different local risks of infection. The larger the total population in the HRC ($N_1=10N_2$) the lower the level of mobility needed to bring the global basic reproductive number below one. This is less sensitive to Community $2$ infection risk, with EVD outbreak control possible as long as the local threshold $\mathcal{R}_{02}<1$ holds.
Further, as the total population in the safer community increases (for example), $\left(N_1=\frac{1}{10}N_2\right)$, the mobility required to control an EVD epidemic turns out to be sensitive to changes in Community $2$ risk of infection and increases in $\mathcal{R}_{02}$.

Figure \ref{fig:PNAS_cordthresh} (right panel) shows the dynamics of the $t_1^-$ threshold under extreme density ratios. Not surprisingly, the more dense the LRC compared to the HRC, the higher level of mobility required to take the final epidemic size below the no-mobility scenario. Moreover, this is sensitive to the Community $2$ infection risk ($\mathcal{R}_{02}$). In fact, when the total population in the LRC is larger ($N_1=\frac{1}{1,000}N_2$), we see that the Community $2$ sanitary conditions ($\mathcal{R}_{02}$) required to take the total final epidemic size below the \emph{cordon sanitaire} via mobility, converge to the sharp threshold $\mathcal{R}_{02}=1$. On the other hand if $\mathcal{R}_{02}>1$ then there is no $t_1^-$ such that the total epidemic size is lower than the size supported by  \emph{cordon sanitaire} scenario level.
In the case when the total population is mainly aggregated in the HRC ($N_1=1,000 N_2$), the \emph{cordon sanitaire} turns out to be less effective at reducing the final epidemic size at most  mobility levels.

\begin{figure}[tbhp]
\centering
\includegraphics[width=0.8\linewidth]{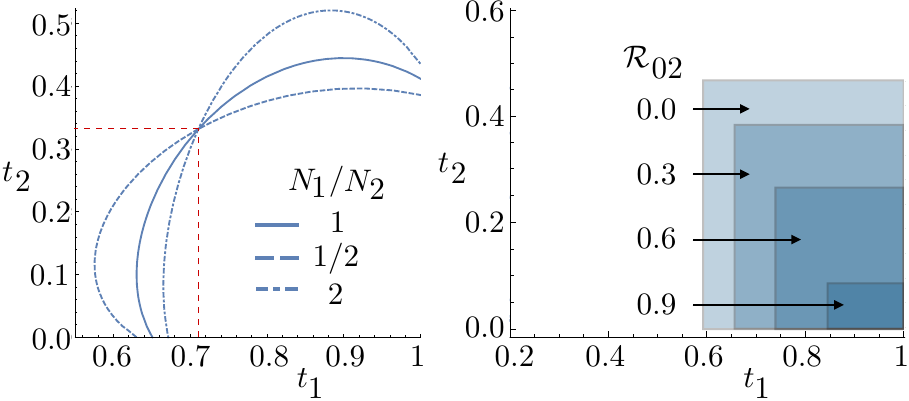}
\caption{(Left panel) Level curves $\mathcal{R}_0(t_1,t_2)=1$, for populations ratios $\frac{N_1}{N_2}=\frac{1}{2},1,2$ and $\mathcal{R}_{02}=0.5$. (Right panel) Mobility regions bringing $\mathcal{R}_0(t_1,t_2)<1$ under extreme population density ratios as function of $\mathcal{R}_{02}$.}
\label{fig:senszone}
\end{figure}

Figure \ref{fig:senszone} (left panel) shows the effects of population density disparities on the threshold condition $\mathcal{R}_0(t_1,t_2)=1$. Simulations suggest that large population size  in the safe community makes the  two-way mobility strategy more effective at reducing the global basic reproductive number below one.
Under the scenario $N_1=\frac{1}{2}N_2$ and $\mathcal{R}_{02}=0.5$, Community $2$ mobility of around  the 10\% level reduces $t_1^+$ below 60\%. On the other hand, Community $2$ null mobility requires a $t_1^+$ level above 60\% in order to attain a global $\mathcal{R}_0$ less than one.

A large population in the HRC ($N_1=2N_2$) allows increases in mobility from the LRC while simultaneously increasing $t_1^+$, complicating the objective of controlling  an EVD outbreak via mobility.

Thresholds in Figure \ref{fig:senszone} (left panel), denoted by $t_1\approx 0.7$ and $t_2\approx0.3$ are used to highlight the minimum Community $1$ and maximum Community $2$ mobility levels capable of bringing the global $\mathcal{R}_0$ below one (for $\mathcal{R}_{02}=0.5$), under extreme population density ratios. In short, there exist a pair of mobility thresholds ``independent'' of population densities, that drive the global basic reproductive number below one. These mobility thresholds are determined by the local basic reproductive numbers, see Figure \ref{fig:senszone} (right panel).

\section{Discussion and concluding remarks}

The problem addressed in this paper -- the role of mobility restrictions in the control of infectious diseases in neighboring communities -- is raised in the most acute way by the imposition of area quarantines or \emph{cordons sanitaires}. By isolating high-risk from low-risk areas, such policies have often made things worse -- increasing rather than reducing the final overall epidemic size. This raises the question of when mobility restrictions are helpful in the management of infectious disease outbreaks. In this paper, we answer that question from a strictly epidemiological perspective. Specifically, we consider when mobility restrictions between two neighboring communities increase the final overall epidemic size, and when they do not.

The present analysis goes beyond  \cite{espinoza2016assessing}, where it was shown that \emph{cordons sanitaires} would not always minimize the total final epidemic size whilst leaving the conditions in which this result held unexplored. Here we find that the lower the relative mobility of people in the high-risk community, the larger the overall final epidemic size; and the lower the relative mobility of people in the low-risk community, the smaller the overall final epidemic size. In the limiting case, when people in the low-risk community are immobile and people in the high-risk community are mobile, allowing unrestricted mobility from the HRC will lead to the elimination of the disease. Our simulations show that limiting the mobility of people in low-risk communities may or may not increase the expected overall final epidemic size, which is a function of the differences in risk. If the low-risk community has a strong enough response to infections, then unrestricted mobility between low- and high-risk communities may reduce and even break transmission chains in the high-risk community. By exporting secondary cases of infection into the low-risk community, the overall production of secondary cases may be reduced.

This aspect of the disease risks of mobility control has not previously been studied.  Of course, there exist scenarios in which the overall production of secondary cases increases with mobility, so increasing the overall final epidemic size. 
Although we find that mobility between high- and low-risk neighborhoods reduces the overall final epidemic size, it increases the low-risk community specific epidemic size, and decreases the high-risk community-specific epidemic size. People moving to the low-risk community generate fewer secondary cases than if they were to remain in the high-risk community, but the total number of secondary cases in the low-risk community goes up.

We also find that the residence time needed by non-residents in the low-risk community to produce a beneficial effect depends on relative population densities in the two areas. If population density is higher in the low-risk community, and if the epidemic cannot be contained in that community ($\mathcal{R}_{02} > 1$), the use of a \emph{cordon sanitaire} strategy can be effective. On the other hand, if population density is higher in the high-risk community, then movement from the high-risk to the low-risk community is likely to reduce the final overall epidemic size. The local risks of infection ($\mathcal{R}_{0i}$) implicitly define a set of mobility thresholds beyond which an epidemic is mathematically not sustainable, regardless of relative population densities.

The important result here is that mobility restrictions may not be an effective policy for controlling the spread of an infectious disease if it is assessed by the overall final epidemic size. Patterns of mobility established through the independent mobility and trade decisions of people in both communities may be sufficient to contain epidemics. For the particular case considered here -- where the two communities are distinguished by health care systems that lead to differences in the level of infection risk -- an increase in the mobility of people residing in the high-risk area may lead to epidemics of shorter duration and smaller size. Since this is the natural response of people facing infectious disease risk, it is worth considering why mobility restrictions up to and including area quarantine are so common.
One explanation may be that low-risk communities place a greater weight on containing risk to themselves than on reducing risk overall. That is, the criterion by which they judge the effectiveness of a disease control policy is not the overall final epidemic size, but the community-specific epidemic size. If the high- and low-risk communities are differentiated by jurisdiction, ethnicity, culture, income, and wealth in addition to the quality of health care, they may be less likely to weight risk-reduction the same in both communities. Our findings abstract from differences in the weights attaching to community-specific disease risk. We suppose that there is a single health authority whose aim is to minimize overall disease risk. But if there are multiple health authorities, each representing a different community, or a different jurisdiction, this is not realistic. Nor is it realistic if there is a single health authority, but it is more responsive to one community than another.

The evidence suggests that at least across national boundaries epidemics are addressed from the perspective of area-specific risk. During the 2009 influenza pandemic, for example, the limited supplies of vaccine where delivered primarily to Canada and the U.S., and not to Mexico. In other cases the international response has been aimed less at reducing the final epidemic size than at containing the disease in high-risk areas -- resource scarce regions, frequently lacking the capacity to manage the isolation of secondary cases. Improving the capacity of such areas to manage disease outbreaks may be the only viable long-term strategy for reducing risk, but is extremely difficult to implement during the course of an epidemic. 

Finally, making informed decisions on the efficacy of mobility restrictions up to and including \emph{cordons sanitaires} depends on our ability to estimate infection risk in different areas. Where we are able to assess risk at different areas \cite{ebola_vac_PNAS}, a Lagrangian approach to the analysis of mobility between them can lead to more effective use of mobility restrictions.

\section*{acknowledgements}
Baltazar Espinoza and Carlos Castillo-Chavez were funded by the National Security Agency (NSA – Grant H98230-J8-1-0005) and partially supported by Data Science Initiative at Brown. Charles Perrings was funded by NSF grant 1414374 as part of the joint NSF-NIH-USDA Ecology and Evolution of Infectious Diseases program, and by UK Biotechnology and Biological Sciences Research Council grant BB/M008894/1.

\bibliographystyle{unsrt}
\bibliography{cordon_unf.bib}

\end{document}